\newcommand{\bea}{\begin{eqnarray}}
\newcommand{\eea}{\end{eqnarray}}

\newcommand{\bt}[1]{{\bar t}}

\documentclass[twocolumn,showpacs,preprintnumbers,amsmath,amssymb]{revtex4}

\usepackage{graphicx}
\usepackage{dcolumn}
\usepackage{bm}

\begin{document}

\preprint{}

\title{\textit{Ab initio} Studies of the Possible Magnetism in
BN Sheet by \\ Non-magnetic Impurities and Vacancies }

\author{Ru-Fen Liu}
 \email{fmliu@phys.ncku.edu.tw}
\author{Ching Cheng}%
 \email{ccheng@phys.ncku.edu.tw}
\affiliation{%
Department of Physics, National Cheng Kung University, 70101, Tainan, Taiwan }%

\begin{abstract}
We performed first-principles calculations to investigate the
possible magnetism induced by the different concentrations of
non-magnetic impurities and vacancies in BN sheet. The atoms of
$Be$, $B$, $C$, $N$, $O$, $Al$ and $Si$ are used to replace either
$B$ or $N$ in the systems as impurities. We discussed the changes
in density of states as well as the extent of the spatial
distributions of the defect states, the possible formation of
magnetic moments, the magnitude of the magnetization energies and
finally the exchange energies due to the presence of these
defects. It is shown that the magnetization energies tend to
increase as the concentrations of the defects decreases in most of
the defect systems which implies a definite preference of finite
magnetic moments. The calculated exchange energies are in general
tiny but not completely insignificant for two of the studied
defect systems, i.e. one with $O$ impurities for $N$ and the other
with $B$ vacancies.

\end{abstract}


\maketitle
\section{Introduction}
The magnetism involving only $s$- and $p$-electron elements
continues to attract much attention due to the potential of
extensive applications as well as the urges to understand its
physical origins. Recently, some experimental groups have
discovered either weak or strong ferromagnetism in
fullerenes\cite{magC,disM} and graphite\cite{graphite} systems. A
few theoretical studies attempting to find magnetism in some
potential non-magnetic systems have also been carried out
previously\cite{possibleM}. However, the origin of ferromagnetism
in those systems is still under debate on both experimental and
theoretical sides\cite{magC,thM}. The mechanism for forming
magnetic ordering in solids, such as ferromagnetism,
antiferromagnetism, etc., can be referred to the direct or
indirect exchange interactions(super-exchange, double exchange,
etc.) among magnetic moments. Therefore, a strong enough exchange
interaction should be the crucial criterion to determine this
possible new class of magnetic materials involving $s$- and
$p$-electron atoms. In this work, the boron nitride(BN) sheet is
used as the host system and the possible formation of magnetic
moments by non-magnetic defects including substitutions of
impurity atoms and creations of vacancies is studied. The exchange
energies of those systems possessing finite magnetic moments are
further examined and we shall demonstrate that there exist mostly
very weak interactions among these defect-induced magnetic moments
except two cases.

BN can form three different bulk structures which are hexagonal
BN(h-BN), cubic BN(c-BN) and wurtzite BN(w-BN). Of these three
structures, h-BN is the room temperature phase. Similar to
graphite, h-BN is quasi-2D with weak interaction between layers.
Nevertheless, different from the delocalized $\pi$ electrons in
graphite, the stronger electronegativity of $N$ than that of $B$
causes $\pi$ electrons to distribute more around $N$. This strong
directional effect of bonding confine the motion of
$\pi$-electrons and thus results in a gap in h-BN. BN is the
lightest III-V compound of those that are isoelectronic with III-V
semiconductors such as $GaAs$, but with wider band gap i.e.
$E_g(BN)=4.0\sim 5.8$ eV at room temperature\cite{gap}, compared
with $1.42$ eV of $GaAs$.

A 2D-structural material is usually a good basis for studying the
physical properties of the most interesting nanostructures, such
as nanotube, nano-ribbon etc.. The tubular structure of BN had
been synthesized\cite{BNtube} experimentally. Moreover, it has
been shown\cite{BNsemi} that, in different ways of rolling up the
BN nanotube, the electronic properties are all semiconducting
which is the same as the extended BN sheet. This is very different
from that of graphite and carbon nanotube which can be either
metallic or semiconducting, depending on the way of rolling up.
Therefore, one would expect that the effects of defects in BN
sheet can catch many important physical properties as those would
exist in the BN nanotubes.

In semiconducting and insulating systems, magnetic moments can be
induced by defects\cite{dM,narrowbn,Madhu}. Here, two different
types of defects are used to study the possible formation of
magnetic moments in BN sheet. One is that the atoms of $Be$, $B$,
$C$, $N$, $O$, $Al$ and $Si$ are used to replace either $B$ or $N$
as impurities. The other is to create vacancies by removing either
$B$ or $N$ atoms in BN sheet.

Our plans for this article are as the followings: Sec. II
summarizes the computational methods in this work. The results of
non-spin-polarized calculations are presented in Sec. III. In this
section, the Subsec. A will present the electronic properties of
BN sheet by the analyzed orbital-projected partial density of
states (DOS). The Subsec. B includes the effects on DOS due to
defects, i.e. the location of the defect states in DOS, as well as
the extent of the spatial distribution of these defect states. All
the results of spin-polarized calculations are presented in Sec.
IV. In the Subsec. A, we will investigate the possible formation
of finite magnetic moments induced by defects while the results of
exchange energies and the corresponding Curie temperatures will be
presented in the Subsec. B. Finally, the conclusions are in Sec.
V.

\section{Computation method}
\begin{figure}
\includegraphics{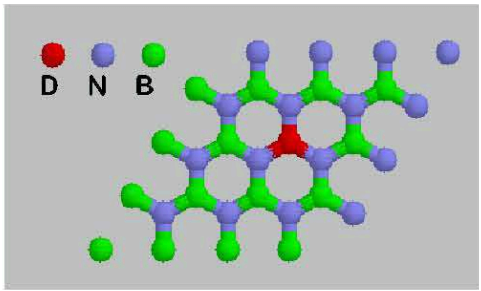}
\caption{\label{bnsheet} (Color online) The $4\times 4$ unit cell
for the hexagonal BN sheet. $D$ represents defects created in the
system, i.e. impurities or vacancies.}
\end{figure}

Our calculations are based on density functional
theory(DFT)\cite{Hohenber} with generalized gradient
approximation(GGA) of Perdew and Wang\cite{PW} for the
exchange-correlation energy functional. The PAW method\cite{PAW}
implemented by Kresse and Joubert\cite{pawimp} is used to describe
the core-valence electrons interactions. There are 2, 3, 4, 5, 6,
3 and 4 valence electrons included here for $Be$, $B$, $C$, $N$,
$O$, $Al$ and $Si$ respectively. The one-electron Kohn-Sham
wavefunctions are expanded by the plane-wave basis with kinetic
energy cut-off ($E_{cut}$ hereafter) of 400 eV. The sampled
k-points in the Brillouin zone(BZ) are generated by the
Monkhorst-Pack technique\cite{Monk} with gamma centered grids for
hexagonal lattice. All calculations are carried out by Vienna
\textit{ab initio} simulation package(VASP)\cite{vasp}.

The calculated lattice constants of h-BN are $a=2.5\AA$ and
$c=6.36\AA$ with $13\times 13\times 5$ k-point grids and compared
with the experimental results($a=2.5\AA$ and $c=6.66
\AA$)\cite{bnbulk} are reasonably well\cite{gga}. The supercells
composing of $4\times 4$ primitive unit cells of BN sheet are used
to simulate systems with defects(See FIG. \ref{bnsheet}). The
vacuum distance between BN sheets (taken as along the $z$
direction) was chosen to be around 13$\AA$ which was tested with
$3\times 3\times 1$ k-point grids to be large enough to avoid
interactions between the sheets.

\begin{figure}
\includegraphics{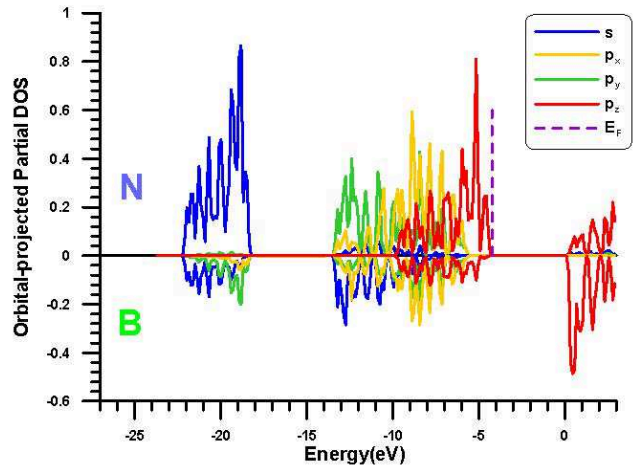}
\caption{\label{bndos} (Color online) The orbital-projected
partial density of states for $N$ and $B$ atoms in hexagonal BN
sheet.}
\end{figure}

Of all the relaxed configurations in this work, the atomic forces
calculated by Hellmann-Feynman theorem\cite{force} were smaller
than $0.02$ eV/\AA. In the spin-polarized calculations, we
performed the above atomic relaxations again to ensure the relaxed
configurations. The DOS and magnetic moments of the final relaxed
structures were then evaluated. The larger supercells of $8\times
4$ and $8\times 8$ were used to study the possible finite magnetic
moments induced by defects, as well as the variation of the
magnetization energy (denoted as $E_M$ hereafter) and the exchange
energy $J$ with respect to the distances between defects, i.e.
defect concentrations.

\section{Non-spin-polarized calculations: Density of States}
\subsection{BN sheet}

The orbital-projected partial DOS of $B$ and $N$ in BN sheet are
presented in FIG. \ref{bndos}. The colors of blue, yellow, green
and red in the figures represent the $s$-, $p_x$-, $p_y$- and
$p_z$-orbital-projected partial DOS respectively. The spiky
feature of the DOS is due to the dimensional reduction from 3D
bulk h-BN to 2D planar structure of BN sheet. Two valance bands
(shorted as VB hereafter) and the lowest 3eV range of the
conduction band (shorted as CB hereafter) are included in the
figure for discussion. The DOS's in the same energy range for BN
sheets with defects will be used in the following discussion for
comparison and the bottom of the CB is taken as the zero of the
energy for convenience in discussing the splitting of the
electronic bands from the two VBs as well as the CB due to the
presence of defects. The two VBs in BN sheet are well-separated as
those in the bulk h-BN\cite{bnbulk}. The lower-energy VB (VB1
hereafter) contains 2 electrons: one is from the $s$-electron of
$N$ and the other is from the hybridized $s$- and $p$-electrons
($sp^2$-electrons hereafter) of $B$. The higher-energy VB (VB2
hereafter) contains 6 electrons, i.e. three $sp^2$-electrons and
one $p_z$-electron from $N$ and two $sp^2$-electrons from $B$.
From FIG.\ref{bndos} we see that the nature of the two VBs for $B$
is mainly hybridized $sp^2$-orbitals while that for $N$ is
$s$-like orbitals in VB1 and hybridized $sp^2$-like orbitals in
VB2. Near the top of VB2 and the bottom of CB, the DOS's are
mostly $p_z$-orbital in nature. However, the dominant contribution
switches from the $p_z$-orbital on $N$ atoms near the top of VB2
to the $p_z$-orbital on $B$ atoms near the bottom of CB. That is
the electronic excitation in BN sheet would involves displacement
of electron distribution from the $p_z$-orbital around $N$ atoms
to the $p_z$-orbital around $B$ atoms.

\subsection{Effect of defects}

\begin{figure}
\includegraphics{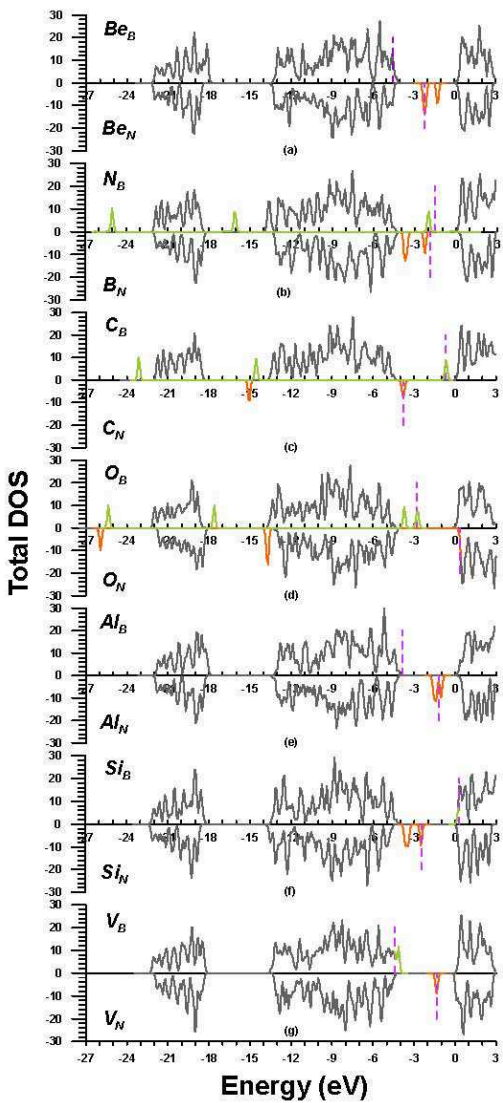}
\caption{\label{aldos} (Color online) The DOS's for the BN sheet
with defects. The upper pannel shows defects on $B$ while the
lower one shows defects on $N$. (a)-(f) are the results of the
defect systems with impurities $Be$, $N$, $B$, $C$, $O$, $Al$ and
$Si$ atoms respectively. The defect systems with $B$ and $N$
vacancies are shown in (g). All dashed lines cut at $E_F$.}
\end{figure}

\begin{figure}
\includegraphics{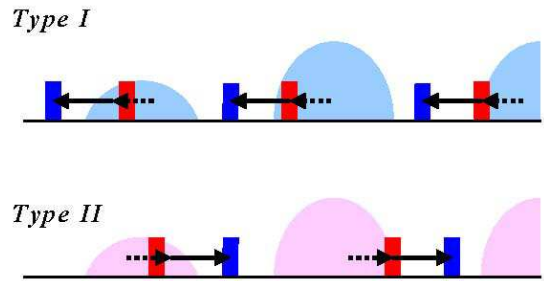}
\caption{\label{type} (Color online) Schematic diagrams of the
effect of defects on DOS. The dashed lines show the possible
circumstances in forming metal-like DOS.}
\end{figure}

In the present section, we discuss the effect of defects on the
DOS of BN sheet when the spin-polarization is not yet switched on.
We shall demonstrate that the effect of defects on the DOS can be
generally categorized into two groups according to the chemical
properties of the defects. The $4\times 4$ supercell was employed
in all the results presented in this section which corresponds to
a doping of 1/32 ($\sim 3\%$) of the defects in the system and a
distance of $10.03\AA$ between the neighboring defects. The
possible interaction between defects will be discussed in the
following section when the exchange energy is investigated.
However, the results presented in the present section are
applicable to the systems with low-density dopings of defects as
we shall demonstrate in the following section that the interaction
between defects is already quite small in the present
configurations. The effect of $C$ impurity on the DOS of BN sheet
will be discussed in details first and then followed by the
results for the series of defects we studied. For simplicity, we
denote, e.g. $C_B$, as the defect system which consists of an
impurity of $C$ atom substituting one of $B$ atoms in the original
supercell of BN sheet and $V_B$($V_N$) as the system consisting of
a vacancy replacing one of the original $B$($N$) atoms.

One of the effects of the impurities is to introduce electronic
states into the energy regions which originally have no DOS in the
BN sheet. We present the total DOS of all the defect systems
considered in this study in FIG.\ref{aldos}. In the case of $C_B$
(FIG.\ref{aldos}(c)), the defect states split from the original
three bands and move to the lower energy region. This is due to
the stronger bonding ability of $C$ ions for electrons compared to
that of $B$ ions. On the contrary, in the case of $C_N$, the
bonding ability of $C$ ion is weaker than that of $N$ ion for
electrons such that the defect bands move to the higher-energy
positions compared to the original three bands. Therefore, the
locations of defect bands is determined by the bonding ability of
the defects compared to the substituted host atoms. The effects of
defects on the DOS's can be summarized into two types as outlined
schematically in FIG.\ref{type}. For $N_B$, $C_B$, $O_B$, $O_N$
and $Si_B$, the bonding abilities of the impurities are stronger
than the substituted host atoms. We name these as type I. For
$Be_B$, $Be_N$, $B_N$, $C_N$, $Al_N$ and $Si_N$, which are named
as type II, the three defect bands all move to the region of
higher energy than the original three bands.

Of these studied defect systems, some consist of odd number of
electrons in the supercell, i.e. $Be_B$, $Be_N$, $C_B$, $C_N$,
$O_B$, $O_N$, $Si_B$, $Si_N$, $V_B$ and $V_N$, such that there is
an unpaired electron occupying the defect state located between
the VB2 and CB for $Be_N$, $C_B$, $C_N$, $O_B$, $Si_N$ and $V_N$,
the top of VB2 for $Be_B$ and $V_B$ as well as the bottom of CB
for $O_N$ and $Si_B$. We denote DB as the state that the unpaired
electron occupies hereafter. We shall show in the next section
that magnetic moments are likely to form for those cases with
partially occupied DBs. In case the partially occupied defect
bands which form definite magnetic moments are also extended in
nature, the long-range magnetic order can develop. Hence, it is
expected that for $Be_B$, $O_N$, $Si_B$ and $V_B$ whose defect
states locating at the edge of the original extended bands, the
defect bands are more likely to be extended in nature and then
more likely to lead to larger exchange energies as will be
presented in the next section.

\begin{figure}
\includegraphics{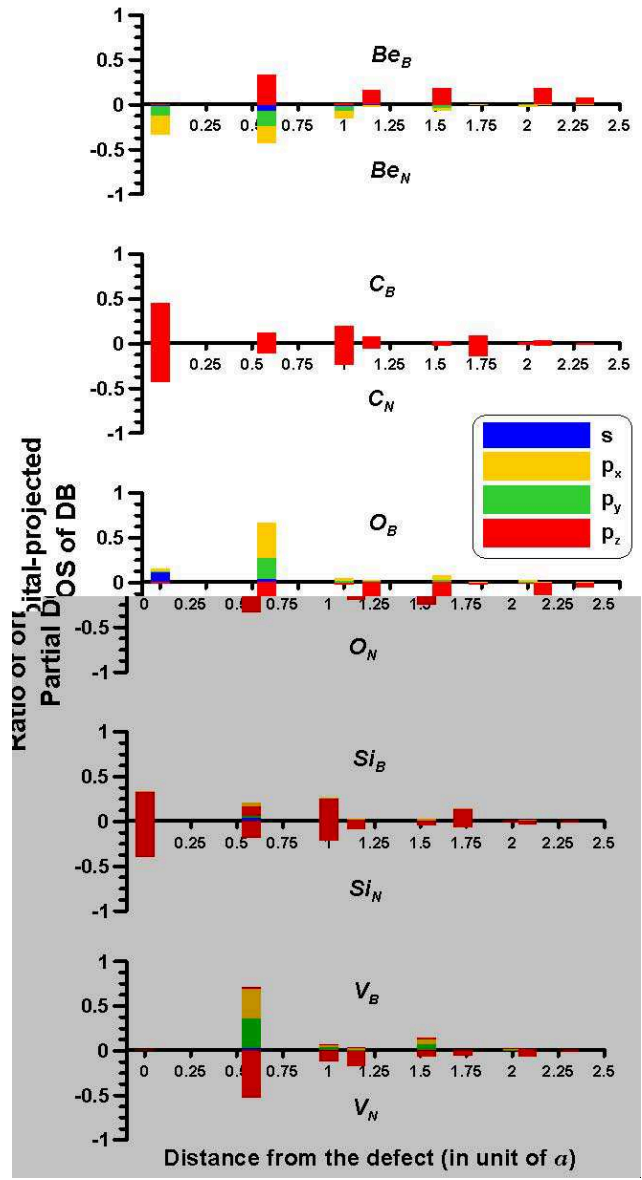}
\caption{\label{peak} (Color online) The ratio of the
orbital-projected partial DOS summing over the $i$th nearest
neighbours of the defect of the system with partially-filled DB.
There are 1(the defect), 3, 6, 3, 6, 6, 3, 3 and 1 atoms
contributing to each bars in sequence. $a$ is the lattice constant
$2.5\AA$.}
\end{figure}

In order to determine the properties of the partially-filled DBs,
the extent of the states in space is established from summing over
the contributions of the orbital-projected partial DOS of the DBs
for the neighboring atoms of the defects as presented in
FIG.\ref{peak}. For $Be_B$, $Be_N$, $O_B$ and $O_N$, we see that
the electrons of the DBs distribute more on the nearest
neighboring atoms of the impurity than the impurity itself. This
is contrary to the $C_B$, $C_N$, $Si_B$ and $Si_N$ systems in
which most of the electrons of the DBs distribute around the
impurity atoms. These results demonstrate that the electrons of
the defect states due to impurities do not have to distribute
mostly on the impurities themselves. About the spatial extent of
the defect states, FIG.\ref{peak} shows that only three systems,
i.e. $Be_B$, $O_N$ and $V_N$, reach beyond $5$\AA . However, the
defect states for all the defect systems considered in the present
study are confined in a distance of $6\AA$ around the defects no
matter whether the partially-filled defect states are at the edge
of the extended bands or not. Finally, we should emphasize that
FIG.\ref{peak} also reveals two categories of DBs: one is formed
by the $sp^2$-hybridized orbitals, e.g. $s$, $p_x$ and $p_y$, and
the other is $p_z$ in nature. These orbital-projected-DOS analyses
imply the difference of the electronic distributions of the
magnetic moments (detail discussions in the next section) formed
by the unpaired electrons in DB, e.g. the moments are distributed
on the plane of BN sheet for $Be_N$, $O_B$ and $V_B$ while the
moments are distributed perpendicularly to the plane of BN sheet
for $Be_B$, $C_B$, $C_N$, $O_N$, $Si_B$, $Si_N$ and $V_N$ systems.

\section{spin-polarized calculations: the effects of defects}

\begin{figure}
\includegraphics{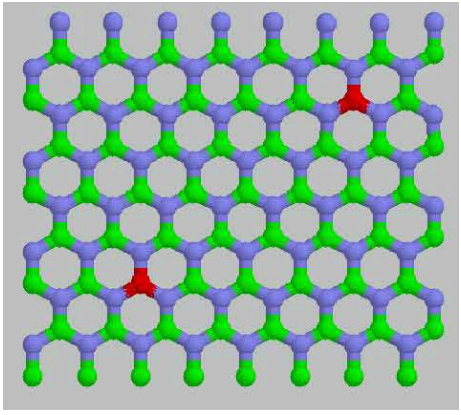}
\caption{\label{841} (Color online) The rectangular supercell with
cell dimensions $20.05\AA$ and $17.37\AA$. The distance between
the two defect sites is $13.26\AA$. The supercell contains 64
primitive cells of BN sheet.}
\end{figure}

\begin{figure}
\includegraphics{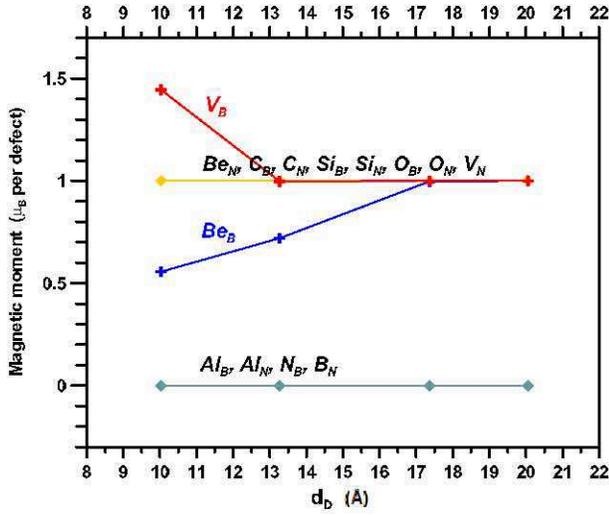}
\caption{\label{mag} (Color online) The calculated magnetic
moments with respect to $d_D$, the distance between the
nearest-neighboring defects. }
\end{figure}

\begin{figure}
\includegraphics{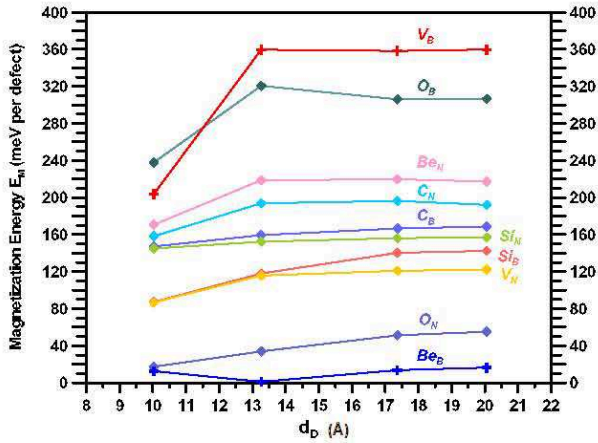}
\caption{\label{em} (Color online) The magnetization energy $E_M$
v.s. $d_D$.}
\end{figure}

\begin{figure}
\includegraphics{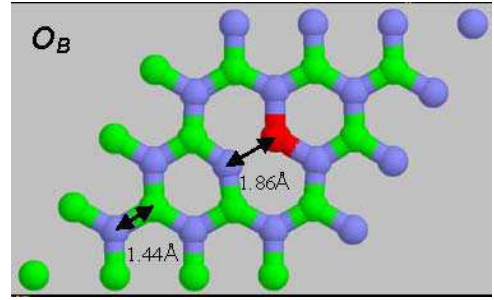}
\caption{\label{ob} (Color online) The relaxed structure for the
defect system of $O_B$. The impurity, $O$ atom, was relaxed and
moved toward two of the three neighboring $N$ atoms to open up one
of the original $N-O$ bonds. The spin-polarized energy for this
configuration with unequal $N-O$ bond lengths is lower than that
with equal $N-O$ bond lengths by 111meV.}
\end{figure}

In this section, we present the results of spin-polarized
calculations to investigate the possible formation of finite
magnetic moment in the BN sheets containing defects as well as the
possible long-range magnetic ordering in these systems.

\subsection{The possible formation of magnetic moments}

To identify the possible magnetic moments due to defects as well
as the extent of the magnetic moments, the $4\times 4$ and
$8\times 8$ supercells consisting of only one defect were employed
to simulate the BN sheet with defect concentrations of $3.125\%$
and $0.78\%$ respectively, i.e. corresponding to a distance of
10.03$\AA$ and 20.05$\AA$ between nearest-neighbour defects
(denoted by $d_D$ hereafter). In order to obtain the relevant
physical quantities for the cases within the range between
$d_D=10\AA\sim 20\AA$, we create a rectangular supercell as shown
in FIG.\ref{841}. By providing either one defect or both defects
in the supercell of FIG.\ref{841}, one can simulate the cases of
17.37$\AA$ and 13.26$\AA$ for $d_D$ respectively. The calculated
magnetic moments of the systems with respect to $d_D$ are
presented in FIG.\ref{mag}. For $N_B$, $B_N$, $Al_B$ and $Al_N$
which have even number of electrons in the supercell and
completely-filled DBs, there exist no magnetizations and they are
therefore excluded from the following discussions. The calculated
magnetic moments in $Be_B$ and $V_B$ were found to vary as $d_D$
increases. However, their magnitudes converge to $1\mu_B$ as $d_D$
increases to 17.37$\AA$. For $Be_N$, $C_B$, $C_N$, $O_B$, $O_N$,
$Si_B$, $Si_N$ and $V_N$, the calculated magnetic moments are
always found to be $1\mu_B$ and independent of $d_D$. Note that
for all the defect systems possessing finite magnetic moments, the
defect bands are always partially occupied. The 1$\mu_B$ of
magnetizations in these defect systems should be attributed to the
unpaired electron occupying the defect states.

To understand how stable these finite moments in BN sheet are, the
magnetization energies, i.e. $E_M$, obtained from the total-energy
difference between the systems with and without spin-polarized
configuration were calculated. The results are plotted in
FIG.\ref{em}. Firstly, the $E_M$s of these systems tend to
increase as $d_D$s increase. This suggests a definite preference
of finite magnetic moments for these defect systems with distant
non-magnetic defects. Secondly, the corresponding $E_M$ for these
defect systems with distant defects can be determined by the
saturated values in the figure, e.g. 360meV for $V_B$, 300meV for
$O_B$, 220meV for $Be_N$ etc., and then are found to be in the
wide energy distribution (from 16meV of $Be_B$ to 360meV of
$V_B$). We notice that the defect systems with metal-like
electronic properties ($Be_B$, $O_N$, $Si_B$ and $V_B$), i.e. with
fermi level at the edge of either VB2 or CB, do not necessarily
lead to large $E_M$. For example, the $E_M$ of $Be_B$ is very
small, i.e. 16meV which is even smaller than the room temperature
thermal energy of 25meV. However, the $E_M$ of $V_B$ is the
largest, i.e. 360meV, of all the systems we considered. We should
mention that, when the defect distance is increased, the
metal-like DOS does not always hold. The bandwidth of the defect
bands might become smaller due to localization such that they
eventually separate from the original VB2 or CB bands. Of the
studied systems with metal-like DOS at $d_D=10.03\AA$, this
happens in $Si_B$ and $V_B$ whose defect bands are originally at
the edge of CB and VB2 respectively. For $Be_B$ and $O_N$, the
defect bands remain at the edge of VB2 and CB respectively up to
the largest $d_D=20.05\AA$ we studied.

Finally, for those systems with $p_z$-like magnetic moments, the
bond lengths between defect and neighboring host atoms were found
relaxed but preserving the three-fold symmetry. However, we found
that, the relaxed bond lengths of $Be_N$, $O_B$ and $V_B$ which
have planar distributions in the magnetic moments(the
$sp^2$-hybridized electron in DB) undergo structural distortions
and break the three fold symmetry of the original BN systems after
the spin-polarization calculation is switched on (John-Teller(JT)
effect). FIG.\ref{ob} shows the relaxed structure of $O_B$. The
impurity, $O$ atom, moved toward two of the neighboring $N$ atoms
to open up one of the three original $N-O$ bonds. The total energy
of this distorted structure with spin-polarization is lower than
that of the structure with equal $N-O$ bond length by 111meV,
similarly for $Be_N$ by 33meV and for $V_B$ by 47meV. The effect
of JT distortion are also responsible for the consequences of
relatively larger magnetization energy of $Be_N$, $O_B$ and $V_B$
with planarly distributed moments compared to those with
$p_z$-like moments.

\subsection{The exchange energy}

\begin{figure}
\includegraphics{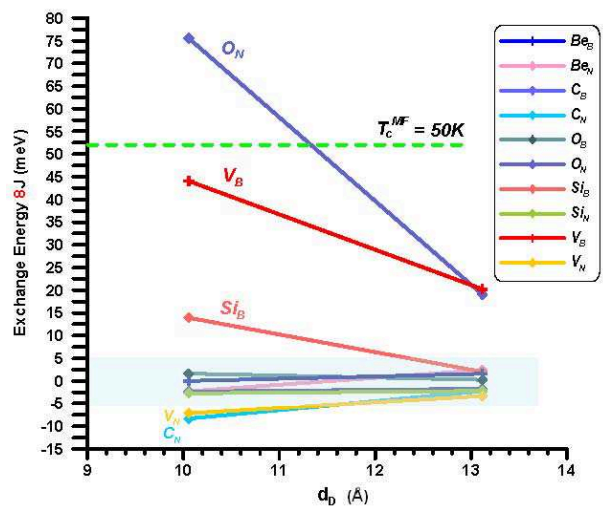}
\caption{\label{8J} (Color online) The exchange energy $8J$ v.s.
defect distances where $J\equiv \frac{1}{8}\{E_{AFM}-E_{FM}\}$.
The energy range from $-$5meV to 5meV is the estimated region
where the magnitude of $8J$ is too small to be used to identify
the sign of $8J$ in our calculations. The dashed line indicates
the value of exchange energy respected to $50K$ of Curie
temperature ($T_c^{MF}$) under mean-field approximation.}
\end{figure}

From the above analyses, we have demonstrated that the
non-magnetic defects can induce finite magnetic moments in BN
sheet. To identify the possible long-range magnetic ordering in
these systems, the Heisenberg-type of spin couplings\cite{J} is
employed to model the interaction ($J$, i.e. the exchange energy)
between the nearest-neighbor magnetic moments due to defects. The
exchange energy $J$ is determined from the total-energy difference
between two spin configurations, i.e. one with antiparallel and
the other with parallel spin configuration whose energy are
denoted as $E_{AFM}$ and $E_{FM}$ respectively hereafter. For
simulating two defects distancing 10.03$\AA$, we employed either
the $8\times 4$ supercell with two defect sites in each
sub-supercells of $4\times 4$ or the $8\times 8$ supercell with
those in the two diagonal sub-supercells of $4\times 4$ to
generate different spin configurations. In the antiferromagnetic
configuration, the $8\times 4$ supercell describes a configuration
of six nearest neighbours ($d_D=10.03\AA$) in which four are
antiparallel- and two are parallel-spin to the centered defect
while the $8\times 8$ supercell describes a configuration of two
antiparallel-spin nearest neighbours ($d_D=10.03\AA$) and a
$d_D=17.37\AA$ for the four next-nearest neighbours. The
rectangular supercell in FIG.\ref{841} with defects on both sites
is also used to simulate the two spin configurations for the
systems with defect distance of 13.26$\AA$. In the rectangular
supercell, there are four nearest neighbours for a defect and all
are antiparallel-spin neighbours in the antiferromagnetic
configuration. We have used both $8\times 4$ and $8\times 8$
supercells for the $V_B$ and $O_N$ systems and the exchange energy
obtained from the $8\times 8$ supercells were considered in the
discussion as they correspond more to the low-doping conditions we
would like to consider. The different values of the exchange
energies obtained from using $8\times 4$ and $8\times 8$
supercells indicate that the simple pair-wise Ising interaction is
not a good enough model to describe the energies of these systems
of $d_D=10.03\AA$. We notice that both $V_B$ and $O_N$ have DB at
the edge of either CB or VB2. The $8\times 8$ supercells were also
used on a few of the other defect systems, but all lead to similar
results as those obtained from the $8\times 4$ supercells.

Our results are summarized in FIG.\ref{8J} where the lines are to
guide the eyes. From the previous magnetization energy
calculations, our computational accuracy can recognize meaningful
energy difference of 5meV and larger values between
configurations. For those systems of $Be_B$, $Be_N$, $C_B$, $O_B$
and $Si_N$, the calculated $8J$s which are less than 5meV are too
small to determine whether the systems are ferro- ($8J>0$) or
antiferro-magnetic ($8J<0$). For $O_N$, $V_B$ and $Si_B$, they are
found to be ferromagnetic, and for $C_N$ and $V_N$,
anti-ferromagnetic at $d_D=10.03\AA$. We should mention that the
relaxed spin configurations of all the systems discussed here
remained anti-ferromagnetic at the end of the calculations if with
initial antiparallel-spin configurations. The exchange energies
for $O_N$, $V_B$ and $Si_B$, which possess metal-like DOS, are
relatively larger at $d_D=10.03\AA$, but not for the case of
$Be_B$ whose $E_M$ is tiny anyway. Although the magnetization
energy of $V_B$ can be as large as 200meV at $d_D=10.03\AA$, the
exchange energy is considerably much smaller (5.6meV), similarly
for $Si_B$ whose $E_M\sim 90meV$ while $J$ is smaller than 2meV.
(Note FIG.\ref{8J} shows the data of $8$ times the exchange energy
$J$.) This contrast is much reduced in the case of $O_N$ whose
$E_M\sim 17meV$ and $J\sim 9.4meV$ at $d_D=10.03\AA$. When the
$d_D$ is increased to $13.26$\AA, the magnitude of exchange
energies for all the above systems decrease. At the end, there are
only two systems, i.e. $O_N$ and $V_B$, which have large enough
exchange energies to be identified numerically as the magnetically
ordered systems.

Within the framework of Heisenberg spin model, it is possible to
estimate the Curie temperatures ($T_c$) of these systems. The
mean-field result is given by \bea
k_BT_c^{MF}=\frac{2}{3}\bar{J}_0, \eea where $\bar{J}_0$ is the
on-site exchange parameter reflecting the exchange field created
by all the neighbouring magnetic moments\cite{ex1,ex2}. Note that
the $T_c$ determined by Eq.(1) is regarded as an overestimated
values\cite{overTc}. Since the physics of $T_c$ involves the
disorder nature of the system due to temperature, $\bar{J}_0$
should be related to the energy difference between the local
disorder state and ferromagnetic state of the system\cite{ex2}.
However, the exchange energies obtained here do not include
disorder effect, i.e. $E_{AFM}$ is calculated from ordered
anti-ferromagnetic configuration. Hence, the estimated $T_c$
(denoted as $\tilde{T}_c^{MF}$ hereafter) by just simply
substituting $J$ in FIG.\ref{8J} into the $\bar{J}_0$ in Eq.(1) is
expected as an upper-bound of $T_c^{MF}$ for a given system. For
$O_N$, $V_B$ and $Si_B$ with relatively large $J$s at
$d_D=10.03\AA$, the $\tilde{T}_c^{MF}$s are $72K$, $43K$ and $13K$
respectively. However, they reduce to around $20K$ for $O_N$ and
$V_B$, below $5K$ for $Si_B$ when $d_D$ is increased to
$13.26\AA$. In addition, the $\tilde{T}_c^{MF}$s of the defect
systems with exchange energies smaller than 5meV are all below
$5K$  in different concentrations. Therefore, one can conclude
that the effect of the long-range magnetic ordering is very weak
in all the defect systems considered here except for the defect
systems of $O_N$ and $V_B$. However, our calculations suggest that
the systems of $Be_N$, $C_B$, $C_N$, $O_B$, $O_N$, $Si_B$, $Si_N$,
$V_B$ and $V_N$ all have finite magnetic moments and are therefore
at least paramagnetic.

\section{Conclusions}

Different concentrations of non-magnetic impurities and vacancies
in BN sheet were studied using first-principles methods to
investigate the possible magnetism in these systems involving only
$s$- and $p$-electron elements. We firstly studied the effects of
these defects on DOS and analyzed the characters as well as the
spacial extent of these defect states. The magnetization energies,
possible magnetic moments as well as the exchange energies for
these defect systems in different concentrations were evaluated.
We demonstrated that all the defect systems with partially-filled
defect bands exhibited a definite preference for finite magnetic
moments. The calculated exchange energies for low-density defect
systems are all tiny except for the $O_N$ and $V_B$ whose exchange
energies are not completely insignificant with an estimated $T_c$
of 20K.

\acknowledgements This work was supported by the National Science
Council of Taiwan. Part of the computer resources are provided by
the NCHC (National Center of High-performance Computing). We also
thank the support of NCTS (National Center of Theoretical
Sciences) through the CMR (Computational Material Research) focus
group.


\begin{thebibliography}{99}
\bibitem{gga} 
The lattice contants calculated from LDA with $E_{cut}=400$eV are
$a=2.48\AA$ and $c=5.9\AA$ and become $a=2.5\AA$ and $c=6.23\AA$
when $E_{cut}$ is increased to 500eV. However, the GGA results of
$c$ become 7$\AA$ at $E_{cut}=500$eV. Similar results of
overestimation for $c$ in hexagonal BN from GGA has been
doccumented in \cite{bnbulk,overgga}.
\end{thebibliography}

\end{document}